\documentclass[useAMS,usenatbib]{mn2e}
\usepackage{natbib}
\usepackage{epsfig}
\usepackage{amsbsy}
\usepackage{hyperref}

\providecommand{\adsurl}[1]{\href{#1}{ADS}}

\newcommand{\lya}{Lyman-$\alpha$~}

\newcommand{\be}{\begin{equation}}
\newcommand{\ee}{\end{equation}}
\newcommand{\ba}{\begin{eqnarray}}
\newcommand{\ea}{\end{eqnarray}}
\newcommand{\brr}{\begin{array}}

\newcommand{\err}{\end{array}}
\newcommand{\bc}{\begin{center}}
\newcommand{\ec}{\end{center}}

\DeclareMathAlphabet{\mathsc}{OT1}{cmr}{m}{sc}
\def\testbx{bx}%
\DeclareRobustCommand{\ion}[2]{%
\relax\ifmmode
\ifx\testbx\f@series
{\mathbf{#1\,\mathsc{#2}}}\else
{\mathrm{#1\,\mathsc{#2}}}\fi
\else\textup{#1\,{\mdseries\textsc{#2}}}%
\fi}
\title[On the importance of high redshift intergalactic voids]
{On the importance of high redshift intergalactic voids}

\author[Matteo Viel,  J\"org M. Colberg \& T.-S. Kim]
{Matteo  Viel$^{1,2}$, J\"org M.\ Colberg$^{3,4}$  \& T.-S. Kim$^{5}$\\
$^1$ INAF - Osservatorio Astronomico di Trieste, Via G.B. Tiepolo 11,
I-34131 Trieste, Italy\\
$^2$ INFN/National Institute for Nuclear Physics, Via Valerio 2,
I-34127 Trieste, Italy\\
$^3$ CMU, Physics Department, 5000 Forbes Avenue, Pittsburgh, PA 15213, USA\\
$^4$ UMass, Department of Astronomy, 710 North Pleasant Street, Amherst, MA 01003, USA\\
$^5$ Astrophysikalisches Institut Potsdam, An der Sternwarte 16, D-14482 Potsdam, Germany \\
\\}

\begin{document}
\maketitle
\begin{abstract}
We investigate the properties of one--dimensional flux ``voids''
(connected regions in the flux distribution above the mean flux level)
by comparing hydrodynamical simulations of large cosmological
volumes with a set of observed high--resolution spectra at $z\sim
2$. After addressing the effects of box size and resolution, we study
how the void distribution changes when the most significant
cosmological and astrophysical parameters are varied. We find that the
void distribution in the flux is in excellent agreement with
predictions of the standard $\Lambda$CDM cosmology, which also fits other flux
statistics remarkably well. We then model the relation between 
flux voids and the corresponding one--dimensional gas density field
along the line--of--sight and make a preliminary attempt to connect
the one--dimensional properties of the gas density field to the
three--dimensional dark matter distribution at the same redshift.
This provides a framework that allows statistical interpretations of
the void population at high redshift using observed quasar spectra, and
eventually it will enable linking the void properties of the
high--redshift universe with those at lower redshifts, which are 
better known.
\end{abstract}


\section{Introduction}

Over the past decade, properties of voids have been more widely
investigated, using different observational probes and tracers (mainly
in the local universe, see for example \cite{hoyle}, \cite{rojas},
\cite{goldberg}, \cite{rojas05}, \cite{hoyle05}, \cite{ceccarelli},
\cite{patiri06a}, \cite{patiri06b}, \cite{tikhonov}), and theoretical
-- analytical or numerical -- models in the framework of the $\Lambda$
Cold Dark Matter ($\Lambda$CDM) concordance cosmology
(e.g. \cite{peebles}, \cite{arbabi}, \cite{mathis}, \cite{benson},
\cite{gottlober}, \cite{sheth}, \cite{goldberg04}, \cite{bolejko},
\cite{colberg}, \cite{padilla}, \cite{furlanettopiran}, \cite{hoeft},
\cite{leepark}, \cite{patiri06c}, \cite{shandarin}, \cite{aloisio},
\cite{tully}, \cite{parklee}, \cite{brunino}, \cite{vdw07},
\cite{neyrinck}, \cite{peebles07}).

The emerging picture is encouraging. On the one hand, some results
appear to be somewhat hard to understand, such as the fact that
galaxies observed at the edges of and in voids appear to be a fair
sample of the whole galaxy population (the so-called 'void
phenomenon', \cite{peebles}), the fact that dwarf galaxies are not
found in voids contrary to expectations from numerical simulations
(e.g. \cite{peebles07}, or the cold spot observed in the Cosmic
Microwave Background (CMB) data (\cite{rudnick}, \cite{naselsky}). On
the other hand, other observational results, such as those from the
2dF or SDSS galaxy redshift surveys, from studies of voids in quasar
spectra, or the CMB spectrum (\cite{caldwell}), are in reasonably good
agreement with theoretical predictions from numerical simulations of
the standard $\Lambda$CDM cosmology. Linking the low--redshift
properties of voids to those at higher redshifts offers the
opportunity to constrain the void population over a significant
fraction of cosmic time and to possibly find out more about voids, to
ultimately understand their role in the context of cosmic structure
formation.

Unfortunately, there are very few observables that constrain the
population of voids at high redshift (see, however, \cite{aloisio}). 
Here, we will use the \lya forest as a tracer of voids 
in the high--redshift universe. The \lya forest (e.g. \cite{bi}) has been 
shown to arise from the neutral hydrogen embedded in the mildly 
non--linear density fluctuations around mean density in the 
Intergalactic Medium (IGM), which faithfully traces the underlying 
dark--matter distribution at scales above the Jeans length. It is 
a powerful cosmological tool in the sense that it probes the 
dark matter power spectrum in a range of scales ($1$ to $80\,h^{-1}$\,Mpc
comoving) and redshifts ($z = $2--5.5) not probed by other
observables. However, the observable is not the matter distribution
itself but the flux: a one--dimensional quantity sensitive not
only to cosmological but also to the astrophysical parameters that
describe the IGM.

The so--called fluctuating Gunn--Peterson approximation \citep{gunnpeterson} 
relates the observed flux to the IGM overdensity in a simple manner once 
the effects of peculiar velocities, thermal broadening, and noise 
properties are neglected: 
\begin{equation}
F({\bf{x}},z)=\exp[-A(z)(1+\delta_{IGM}({\bf{x}},z))^{\beta}]\, ,
\end{equation} 
with $\beta(z) \sim 2-0.7(\gamma(z)-1)$, $\gamma$ the parameter 
describing the power--law ('equation of state') relation
of the IGM, $T=T_0(z)((1+\delta_{IGM})^{\gamma-1}$, and $A(z)$ a
constant of order unity which depends on the mean flux level, on the
Ultra Violet Background (UVB) and on the IGM temperature at mean 
density. In practice, linking the flux to the density of the dark 
(or gaseous) matter requires the use of accurate hydrodynamical 
simulations that incorporate the relevant physical and dynamical 
processes. However, for approximate calculations equation\,(1), 
in conjunction with simple physical prescriptions of \lya clouds 
(e.g. \cite{schaye}), can offer precious insights. 

Provided the flux--density relation is accurately modelled, using the
\lya forest to check the properties of the population of voids in the
matter distribution would avoid having to know the bias between
galaxies and dark matter and would allow to sample over larger volumes
the values of cosmological parameters that affect the void population. 
On the other hand, the fact that the flux information is
one--dimensional requires non--trivial efforts to relate it to the
three--dimensional (dark) matter density field.  Motivated by huge
amounts of new data of exquisite quality, we here present a first
attempt in this direction. Note that in the early 1990s, before the
new paradigm of the \lya forest absorption was proposed, voids from
the transmitted \lya flux of QSO spectra were analysed, using discrete
statistics on the lines, by several groups
(e.g. \cite{crotts,duncan,ostriker,dobrzycki,rauch92}). These
attempts, however, did not reach conclusive results due to
small--number statistics. We defer to a future paper for a link
between the observational properties outlined here and those at low
redshift (e.g. \cite{mathis,hoyle}) and for a careful investigation of
the absorber--galaxy relation (e.g. \cite{stocke,mclin}).

This paper is organized as follows. In Section \ref{data}, we briefly
describe the data set used. In Section \ref{simulations} we
present the hydrodynamical simulations. Section \ref{results}
contains the bulk of our analysis and the results, which are
summarized in the conclusions in Section \ref{conclusions}.

\section{The data set}
\label{data}
We use the set of 18 high signal--to--noise (S/N$>30$--$50$),
high--resolution ($R\sim 45000$) quasar spectra taken with the
VLT/UVES presented in \citep{tkim}. We regard these spectra as a
state--of--the--art, high--resolution sample in the sense that many
statistical properties of the transmitted flux (mean flux decrement,
flux probability distribution function, and flux power and bispectrum)
have been accurately investigated and interpreted by several authors
from the same or from similar data sets (see, for example,
\cite{mcdonald00}, \cite{kim04}, \cite{jena}, \cite{vielbispect},
\cite{bolton07}). More importantly, systematics effects such as metal
contamination, noise and resolution properties, and continuum fitting
errors have been extensively addressed. From the data set, we discard
the four highest redshift QSOs, in order to have a more homogenous
sample at a median $<z> = 2.2$ and with a total redshift path $\Delta
z=6.1$.  The pixels contributing to the signal are in the wavelength
range $3282$--$4643\, \AA$ while the wavelength ranges for each single QSO
spectrum are reported in \cite{tkim}.  Both Damped \lya systems and
metal lines have been removed from this sample.  The evolution of the
effective optical depth derived from this data set, $\tau_{\rm
  eff}=-\ln <F>$, which -- as we will see -- is the main input needed
to constrain and identify voids, has been widely discussed and
compared with various other measurements \citep{tkim}.

\begin{figure*}
\begin{center}
\includegraphics[width=16cm, height=16cm]{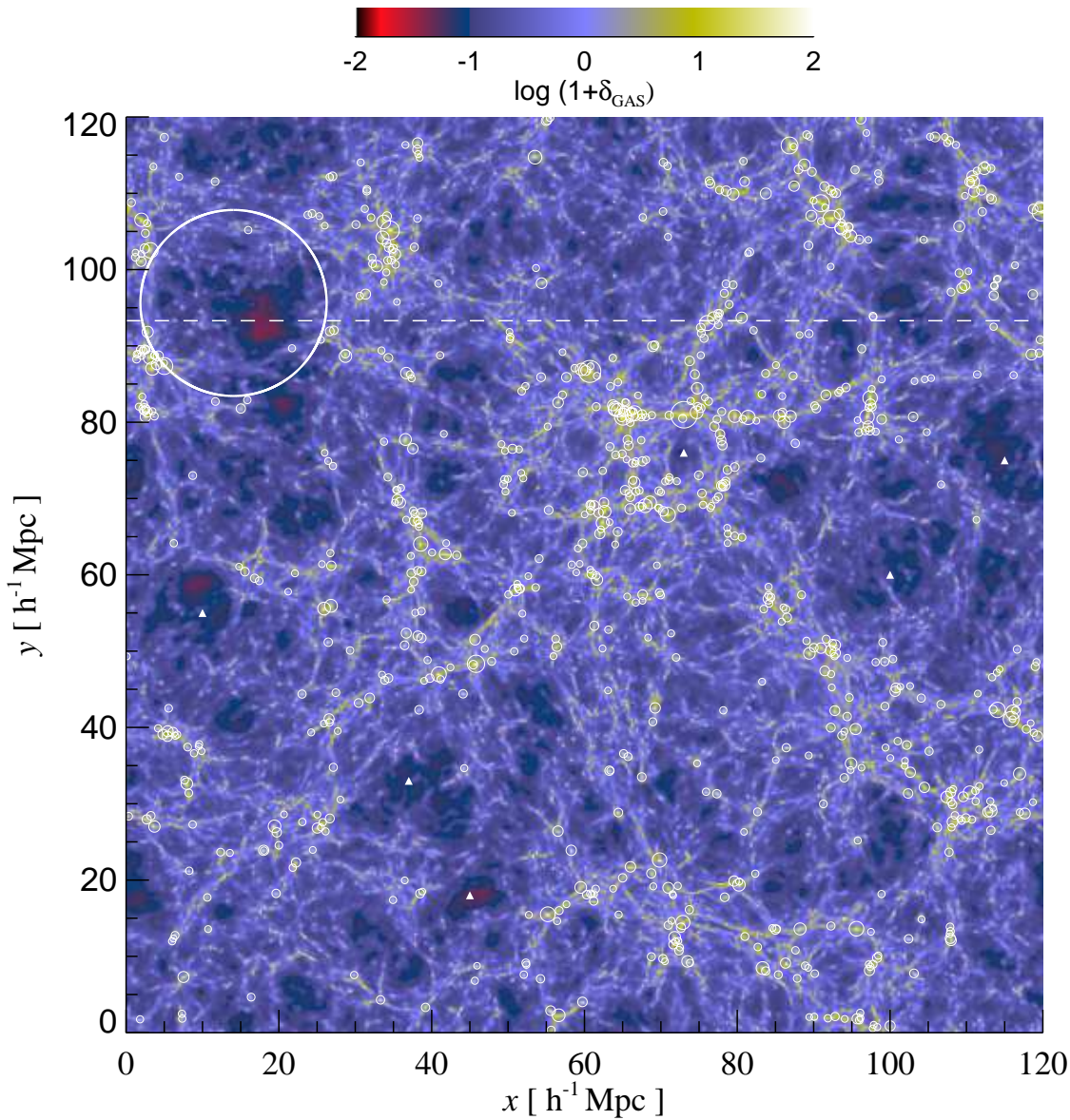}
\end{center}
\caption{Slice of the projected gas density (thickness is 10 Mpc$/h$)
  from our reference $\Lambda$CDM hydrodynamical simulation at
  $z=2.2$. The presence of roughly spherical regions below the mean
  density of sizes of the order tens of Mpc is clear, denser
  environments at $\delta \sim 100$ tend to line up in filaments or
  could be found (more rarely) inside underdense regions (voids). A
  large underdense region is shown as a white thick circles, while
  haloes with total masses above $2\times10^{11} M_{\odot}$, that
  trace the cosmic web, are represented by smaller white circles with
  radii proportional to their masses. The dashed line--of--sight will be
  analysed more closely in Section \ref{3d}. Filled triangles indicate
  some underdense regions of different sizes: usually smaller voids
  are surrounded by many massive haloes, while larger ones have less
  haloes at their edges (with smaller masses).}
\label{slice}
\end{figure*}

\section{The simulations}
\label{simulations}
We use a set of simulations run with {\small {GADGET-2}}, a parallel
tree Smoothed Particle Hydrodynamics (SPH) code that is based on the
conservative `entropy-formulation' of SPH \citep{springel}.  The
simulations cover a cosmological volume (with periodic boundary
conditions) filled with an equal number of dark matter and gas
particles.  Radiative cooling and heating processes were followed for
a primordial mix of hydrogen and helium. We assume a mean UVB produced
by quasars and galaxies as given by \cite{haardt1996}, with helium
heating rates multiplied by a factor $3.3$ in order to better fit
observational constraints on the temperature evolution of the IGM.
This background naturally gives $\Gamma_{12}\sim 1$ at the redshifts
of interest here \citep{bolt05}.  The star formation criterion very
simply converts all gas particles whose temperature falls below $10^5$
K and whose density contrast is larger than 1000 into (collisionless)
star particles (it has been shown that the star formation criterion
has a negligible impact on flux statistics, see for example
\cite{viel04} and \cite{bolton07}). More details of the simulations
can be found in \citep{viel04}.

The cosmological model corresponds to the `fiducial' $\Lambda$CDM
Universe with $\Omega_{\rm m
}=0.26,\ \Omega_{\Lambda}=0.74,\ \Omega_{\rm b}=0.0463$, $n_s=0.95$,
and $H_0 = 72$ km s$^{-1}$ Mpc$^{-1}$ and $\sigma_8=0.85$ (the B2
series of \cite{viel04}). We use $2\times 400^3$ dark matter and gas
particles in a volume of size $120\ h^{-1}$\,Mpc box. For cross checks
we also analyse smaller boxes of size $60, 30,$ and $15\,h^{-1}$\,Mpc.
The gravitational softening was set to $15\,h^{-1}$\,kpc in comoving
units for all particles for the largest simulated volume and changes
accordingly for the smaller volumes. In the following, the different
simulations will be referred to by the tuple (comoving box size in
Mpc$/h$,(no. gas particles)$^{1/3}$), so (120,400) denotes our
fiducial simulation etc. (all the sizes reported in the rest of the
paper will be in comoving units).  In Figure \ref{slice} we show a
two--dimensional slice of thickness 10 comoving Mpc$/h$ of the gas
density for our reference simulation at $z=2.2$. Large regions whose
gas density is below the mean are present and overdense regions
separate them and give rise to the cosmic web. Occasionally denser
regions of overdensity $\sim 100$ reside in the inner regions of
voids, however their sizes are quite small and their impact parameter
in the lines--of--sight to distant QSOs should be also relatively
small. In Figure \ref{slice} we also overplot the haloes whose total
mass is above $2\times 10^{11} M_{\odot}$ (white circles) and show
some underdense regions of different sizes (filled
triangles). Usually, the larger voids are surrounded by few and not
very massive haloes that line up in filaments, while smaller voids,
such as the one at $(x=77,y=73)$ Mpc$/h$ have more spherical shapes
and have more massive haloes around them.

Note that some numerical efforts in order to simulate the intergalactic
voids are required: on one side large volumes need to be simulated in
order to sample the void distribution at large sizes, on the other
side relatively high resolution is mandatory. In fact, with poor
resolution the denser regions at the edge of voids are not properly
resolved and will result in a smaller amount of neutral hydrogen and
thereby less absorption: this will cause an overestimate of the void
sizes.

\begin{figure}
\begin{center}
\includegraphics[width=9cm, height=9cm]{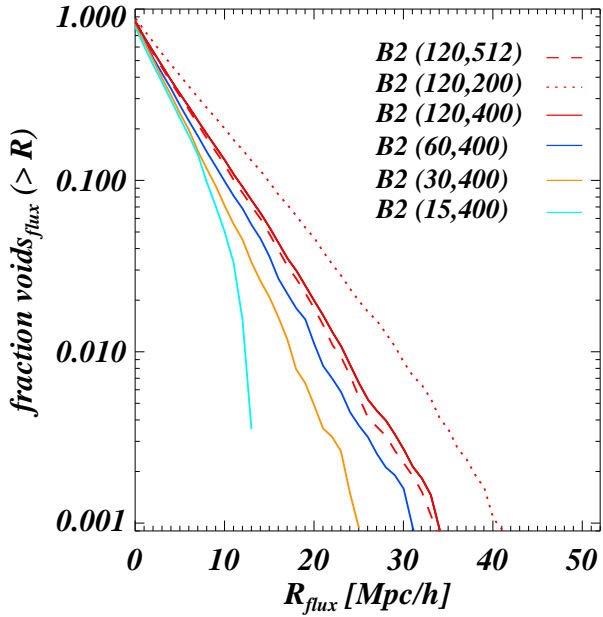}
\end{center}
\caption{Box size and resolution effects at $z=2$ on the fraction of
  one--dimensional voids in the flux distribution (of size $R$ or
  larger), extracted from four different simulation volumes (120, 60,
  30 and 15 $h^{-1}$\,Mpc) and three different resolutions ($200^3,
  400^3, 512^3$ gas particles). The flux has been smoothed with a 1D
  top--hat filter of size $1\,h^{-1}$\,Mpc, which roughly corresponds
  to the Jeans length at this redshift. Spectra have been normalized
  to the same effective optical depth.}
\label{fig1}
\end{figure}

The parameters chosen here, including the thermal history of the IGM,
are in reasonably good agreement with observational constraints and
with recent results from CMB data, weak lensing and other data sets
obtained from the \lya forest (e.g. \citet{viel06}, \citet{lesg}).  The
mock QSO spectra extracted from the simulations have the same
resolution and noise properties as the observed ones, even though this
will not affect our results. Here, for two reasons we consider only
simulation snapshots at $z\sim 2$. First, the high--resolution data
set has a median redshift $<z>=2.2$. Second, at these redshifts the
UVB should be uniform, and the IGM's equation--of--state
is reasonably measured, leaving little room for ionization voids to be
labelled as matter voids or for temperature fluctuations to strongly
affect the \lya forest properties \citep{shang}. However, we caution
that: $i)$ recent results from \cite{bolton07}, based on the flux
probability distribution function, seem to suggest that the thermal
state of the IGM at $z<3$ could be more complex than expected; $ii)$
patchy HeII reionization at $z\sim3.2$ could affect the uniformity of
the UVB (e.g. \cite{bolton06,faucher}) and some effects at smaller
redshifts might influence \lya opacity.

\section{Results}
\label{results}

\subsection{The 1D flux void distribution}
\label{1Dsect}

\begin{figure}
\begin{center}
\includegraphics[width=9cm, height=9cm]{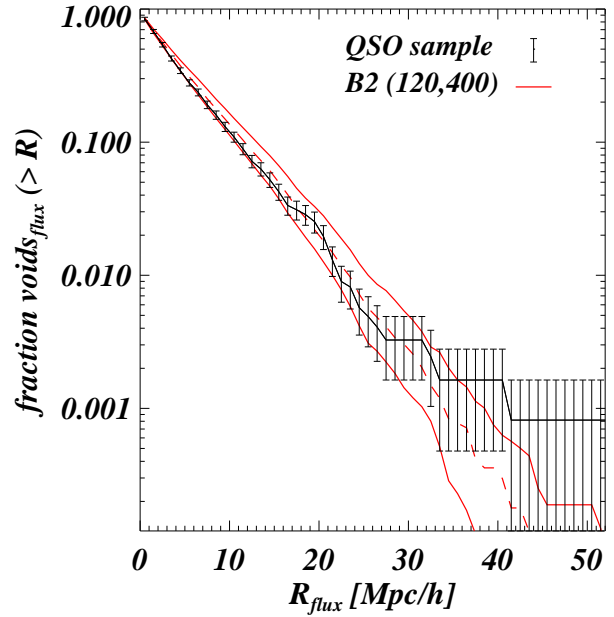}
\end{center}
\caption{Fraction of 1D voids in the flux distribution at $z=2.2$ of
  size larger than $R$ as extracted from the observed sample
  ($1\sigma$ Poissonian error bars). Continuous lines represent the
  reference hydrodynamical simulation ($120$\,Mpc$/h$) with voids
  identified taking $\pm 3\sigma$ values around the observed effective
  optical depth value (dashed line). The flux has been smoothed with a
  1D top--hat filter of size $1$\,Mpc$/h$.}
\label{fig2}
\end{figure}

We search for flux voids in the simulated spectra in the following
simple way: $i)$ we extract 1000 lines--of--sight from the simulated
volume; $ii)$ we normalize the mock spectra to have the observed mean
flux value; $iii)$ we smooth the flux with a top--hat filter of size
$1\,h^{-1}$\,Mpc, which roughly corresponds to the comoving Jeans
length; $iv)$ we identify connected flux regions above the mean flux
level at the redshift of interest. The normalization is performed by
multiplying the whole array of 1000 simulated optical depths arrays by
a given constant (found iteratively after having extracted the spectra)
in order to reproduce the wanted $\tau_{\rm eff}$ value. The choice of
a different mean flux will of course impact on the void selection
criterion, since a smaller (larger) mean flux level will increase
(decrease) the number of pixels at high (low) transmissivity that are
likely to correspond, in a non--trivial way, to under(over)-dense
regions.

As a first preliminary test we check for the effects of resolution on
the flux fractions of voids at $z=2$. Results are shown in
Fig.~\ref{fig1}, which shows simulations with different box sizes and
resolutions, but with the same initial conditions normalized to the
same mean flux. Note that this analysis is performed in redshift space
and peculiar velocities affect the flux properties.  From the Figure,
it is clear that in order to sample the fraction of voids accurately a
large enough volume and relatively high resolution are needed. For
example, the $(120,200)$ simulation differs dramatically from the
$(120,400)$ one, because in the latter relatively dense regions placed
at the edge of void regions are better resolved. However, this
discrepancy disappears once the flux is smoothed over a given scale
(the Jeans length in our case). In a way, smoothing the flux degrades
the resolution of the simulation making it less sensitive to the
relevant physical processes causing absorption at and below the Jeans
scale.  Also note how increasing resolution beyond that in the
$(120,400)$ simulation -- the $(120,512)$ case -- produce results that
are well within the statistical error bars that are observed and will
be shown in Figure \ref{fig2}. Because of this, $(120,400)$ is now
taken to represent our fiducial run that will be used to explore other
physically meaningful parameters.

We now perform the same search for the observed QSO spectra.  The
results are shown in Figure \ref{fig2}, which shows the distribution
of the flux void population extracted from the observations (points
with Poissonian error bars) and from our largest volume
simulation. From the Figure it is clear that the fiducial $(120,400)$
run is in excellent agreement with observations and that flux voids of
sizes $> 35\,h^{-1}$\,Mpc are about 1000 times less common than voids
of sizes larger than few $h^{-1}$\,Mpc. We also note a bump at around
$20$ Mpc/$h$ in the voids fraction, but one would need a larger sample
of QSOs in order to better test its statistical significance.  The
continuous lines show how the void fraction changes when the effective
optical depth is varied at a $\pm 3\sigma$ level around the observed
value. In particular, the lower and higher $\tau_{\rm eff}$ values
($\tau_{\rm eff}=0.136$ and $\tau_{\rm eff}=0.19$, respectively) have
been chosen in such a way to conservatively embrace at a confidence
level of $3\sigma$ the values obtained in \cite{viel04}, where the
measured value for the effective optical depth was $\tau_{\rm
  eff}=0.163\pm0.009$.  The recently determined power-law fit of
\cite{tkim} $\tau_{\rm eff}=(0.0023\pm 0.0007) (1+z)^{3.65\pm 0.21}$
by using QSO spectra in the range $1.7<z<4$, is influenced by the
scatter coming from QSO spectra over a large range of redshift, and
would in principle allow for a larger range of effective optical
depths. However, by selecting only a subsample at a given redshift,
smaller statistical errors on the effective optical depth can be
obtained.

We stress that the definition of void used here is based on the
transmitted flux and thus is very different from studies based on the
distribution of voids as inferred from the galaxy distribution.

\begin{figure}
\begin{center}
\includegraphics[width=9cm, height=9cm]{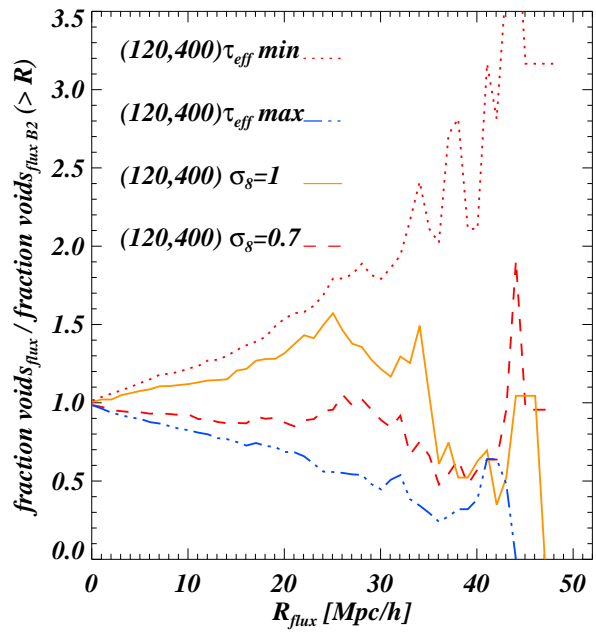}
\end{center}
\caption{Ratio of the flux void fraction of the $(120,400)$ simulation
  ($\sigma_8=0.85$) with runs that use a different value of the
  power--spectrum amplitude ($\sigma_8=0.7$ and $\sigma_8=1$,
  represented by the dashed and continuous lines, respectively). Also
  shown are the ratios when the effective optical depth (which is used
  as a threshold to define voids in the flux distribution) of the
  reference run is conservatively varied between the observational
  bounds ($\pm 3\sigma$ values represented by the dot--dashed and
  dotted lines). Spectra have been normalized
  to the same effective optical depth.}
\label{fig3}
\end{figure}

\subsection{Exploring the parameter space}
\begin{figure}
\begin{center}
\includegraphics[width=9cm, height=9cm]{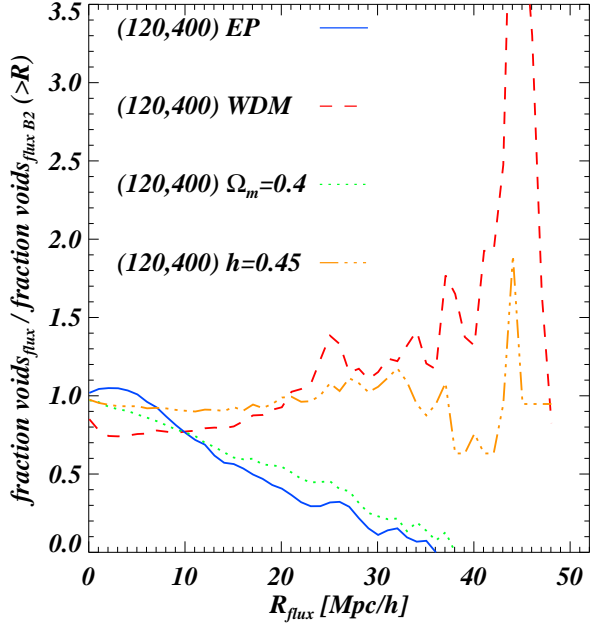}
\end{center}
\caption{Ratio of the flux void fraction of the $(120,400)$ simulation
($\sigma_8=0.85$) with runs that have some extra power (EP, continuous
line) and less power (WDM, dashed line) at Mpc scales. Also shown are
the ratios when the Hubble constant is set to a smaller $h=0.45$ value
(dot-dashed line) and a simulation with a larger value of
$\Omega_m=0.4$ (dotted line). The reference B2 (120,400) run has
$h=0.72$ and $\Omega_m=0.26$. Spectra have been normalized
  to the same effective optical depth.}
\label{fig4}
\end{figure}

We now discuss the main cosmological and astrophysical parameters that
can affect the population of voids in the flux distribution.  In
Figure \ref{fig3}, we plot the ratios of the flux void fractions in
the fiducial and in other simulations, where we changed $\tau_{\rm
  eff}$ at $z=2.2$ within the observational bounds and the value of
$\sigma_8$.  As could be expected, the fraction of voids changes
dramatically with a change in $\tau_{\rm eff}$: at $z\sim 2$ the flux
probability distribution function (pdf) is a steep function around the
mean flux level. Thus adopting a different void selection criterion
has a large impact on the void distribution.  In the following, we
normalize the mock QSO sample to reproduce the same effective optical
depth. Changing $\sigma_8$ appears to have a smaller effect. A
simulation with $\sigma_8=1$ results in a faster evolution of cosmic
structures and in a 50$\%$ increase of the void population of size
$R=20-35\,h^{-1}$\,Mpc, compared with the reference $\sigma_8=0.85$
case.  In the simulation with a lower value of $\sigma_8$, the void
sizes are more evenly distributed. Note, however, that the trend at
the largest sizes (45\,Mpc$/h$) is somewhat different: here the low
$\sigma_8$ simulation predicts more voids than the higher $\sigma_8$,
since the smaller amount of power in the former can occasionally
produce large regions devoid of matter and absorption.  We also
checked the impact of varying other parameters such as $\Omega_m$,
$H_0$ and a change in the shape of the linear dark matter power
spectrum in the initial conditions to account for a presence of warm
dark matter (WDM) and adding some extra power (EP) at intermediate
(from Mpc to tens of Mpc) scales, that results in an effect opposite
to that of WDM. The mass chosen for the WDM particle is 0.15 keV. A
mass that is already ruled out from the flux power spectrum of the
\lya forest (see \cite{viel08}) and that produces a suppression
of power at scales around 0.4 Mpc$/h$, which are marginally sampled in
the initial conditions of our (120,400) simulation. However, for the
qualitative purposes of this paper this simulation should show the
impact of a WDM particle on the distribution of the largest voids in
the flux. At the scale of 2.5 Mpc$/h$ the power is 100 times higher
(smaller) in the EP (WDM) simulation than the corresponding (120,400).
The results are plotted in Figure \ref{fig4}, where all the
simulations have the same value of $\sigma_8=0.85$: it is clear that
the WDM and EP runs have opposite effects.  WDM significantly enhance
by a factor larger than 3 the fraction of largest voids, by erasing
substructures below a given scale, while the EP simulation presents
more collapsed structures that suppress the number of the largest
voids.  The effect of a larger value of $\Omega_m=0.4$ is similar to
that of the EP simulation and determines a 90\% reduction in the
fraction voids whose sizes are larger than 30 Mpc$/h$. Finally, we
checked for the effect of a smaller value of $H_0=45$ km/s/Mpc a value
proposed recently by \cite{stephon} to be the true value of the Hubble
constant outside a local underdense region of very similar density to
those explored here. Having a smaller value for the Hubble constant
produces effects that are very similar to those of a smaller value of
$\sigma_8=0.7$ (see Fig. \ref{fig3}): the evolution of the cosmic web
is less pronounced and can occasionally produce large regions devoid
of absorption, even if on average the effects are quite mild. We also
varied the spectral index $n_s$ but found small differences to
the reference case.  As can be appreciated from Figure \ref{fig2}, the
statistical errors on the observed void fraction are somewhat large
and thereby distinguishing between different cosmological model is
still very difficult and none of the models presented here could be
convincingly ruled out.

\begin{figure}
\begin{center}
\includegraphics[width=9cm, height=9cm]{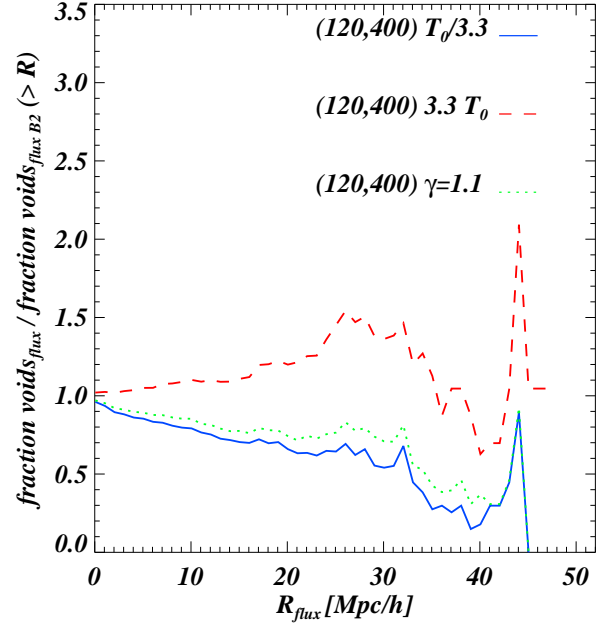}
\end{center}
\caption{Ratio of the flux void fraction for the $(120,400)$ runs,
comparing the fiducial run ($\gamma=1.6$ and $T=T_0$) with runs that 
have a different value of the IGM temperature at mean density 
($2\,T_0$ and $T_0$ are represented by dashed and continuous lines, 
respectively) and with a run that has a different value of $\gamma$ 
(dotted line). Spectra have been normalized
  to the same effective optical depth.}
\label{fig5}
\end{figure}

\begin{figure*}
\includegraphics[width=18cm, height=9cm]{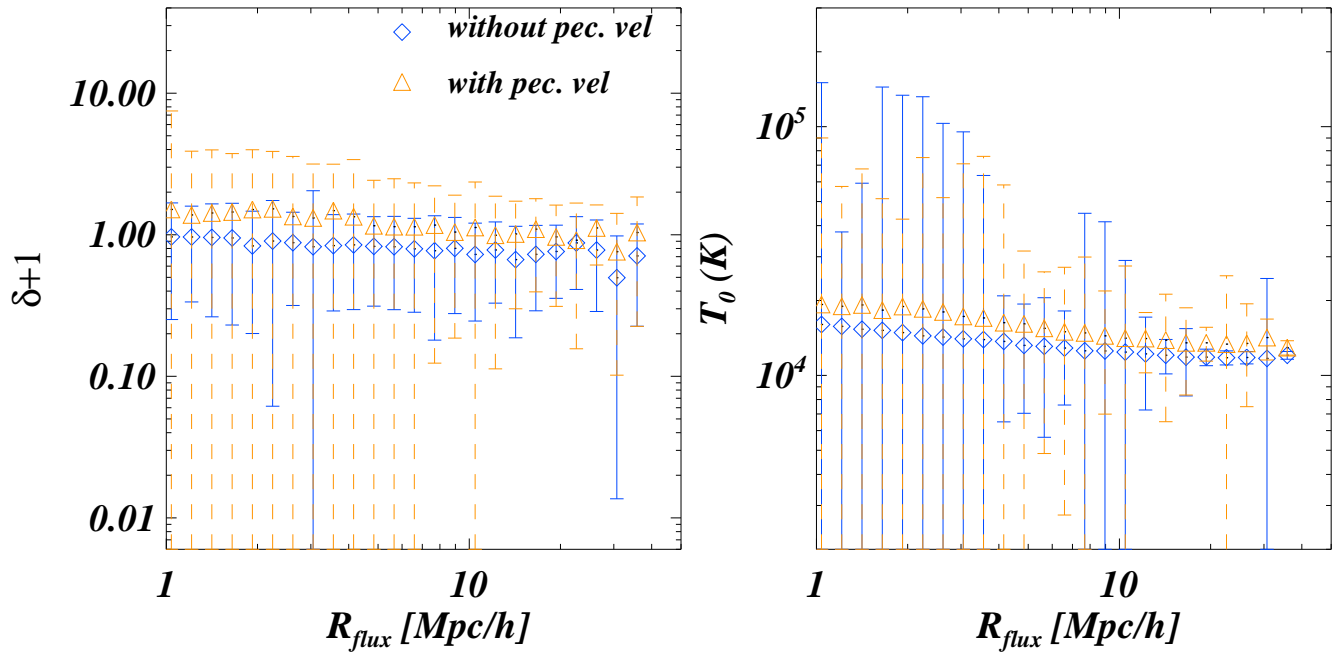}
\caption{1D gas density and temperature in void regions along
lines--of--sight extracted from the fiducial simulation at $z=2.2$. 
Voids have been identified from the flux distribution as regions 
above the mean flux level. Results are shown with (triangles) and without
(diamonds) peculiar velocities, and error bars represent the scatter.}
\label{fig6}
\end{figure*}

Several other parameters could have an impact on the void
distribution, in particular the two parameters that are poorly
constrained and that describe the thermal state of the IGM: $T_0$ and
$\gamma$.  We investigate the effect of a different value of $T_0$ in
Figure~\ref{fig5}. A colder (hotter) IGM will result in a higher
(lower) density of neutral hydrogen and thereby in more absorptions,
which, in turn, will decrease (increase) the fraction of large voids.
In Figure~\ref{fig5}, we also overplot the effect of changing $\gamma$
from the reference value ($\gamma=1.6$) to $\gamma=1.1$ (represented
by the dotted line in the Figure), which is a better fit to the data
at this redshift \citep{bolton07}. A flatter equation of state will
not only make underdense regions hotter, but it will also make the
slightly overdense regions colder. The overall effect is a slight
decrease in the fraction of flux voids, since cold regions embed on average more
neutral hydrogen than hot ones. This result shows that by selecting
mean--flux--level regions we are sampling regions around mean density
(for which the power--law 'equation of state' could be a poor fit to
the actual distribution).

\begin{figure}
\includegraphics[width=9cm, height=9cm]{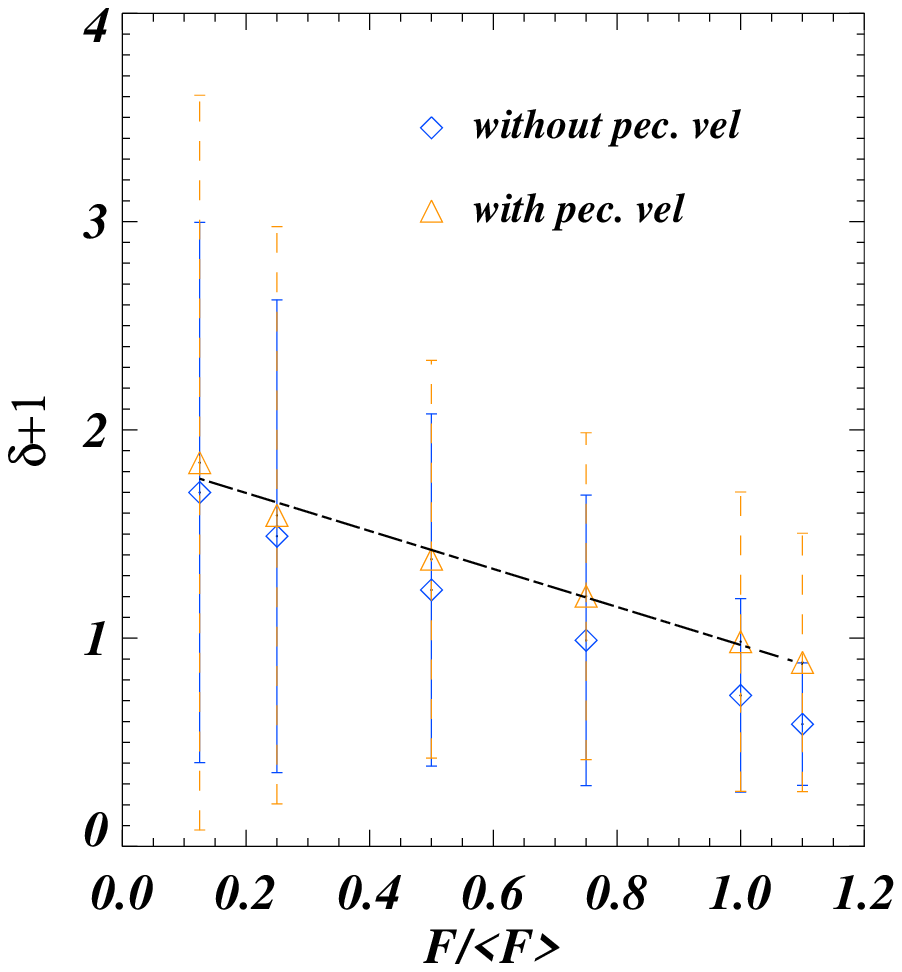}
\caption{Densities of voids (larger than 7$\,h^{-1}$\,Mpc) as a 
function of flux threshold level at $z=2.2$.  Results are
shown with (triangles) and without (diamonds) peculiar velocities, 
and error bars represent the scatter. The dashed line
shows a linear fit to the case with peculiar velocities.}
\label{fig1D_2}
\end{figure}

Note that the effect of changing $\gamma$ was addressed by running a
new simulation that self--consistently results in a low $\gamma$ at
$z=2$ (using a modified version of the \small{GADGET-2} code kindly
provided by J. Bolton) and not by rescaling the gas-temperature
relation a posteriori as is sometimes done.  Another important input
parameter is the overall amplitude of the UVB, parametrized by
$\Gamma_{-12}$, for which we have some observational constraints
\citep{bolt05}. However, this quantity is degenerate with the value of
$\tau_{\rm eff}$, so changing $\Gamma_{-12}$ will result in the same
change as changing $\tau_{\rm eff}$.

\subsection{From flux to 1D gas density}
\label{1Dflux}

Having identified voids at $z\sim 2$, using the mean flux as a
threshold, we are now interested in linking the flux properties to the
underlying physical properties of the gas density field.  The main
problem we have to deal with is that the flux is observed in redshift
space, while we want to recover gas properties in real space. The role
of peculiar velocities in determining the properties of the
transmitted flux has been investigated by several authors
(e.g. \cite{weinberg}), however for the purpose of the analysis
performed here it produces a shift of the absorption feature from the
real space overdensity to the redshift space observed flux. This
increases the scatter since redshift space pixels are not related in a
simple manner to the underlying real space density field.

Occasionally, peculiar velocities could be of the order of a few
hundred km/s, so the correspondence is non-trivial (see for example
\cite{rauchetal}).  However, we can rely on hydrodynamical simulations
in order to infer, at least on a statistical basis, the physical
properties of the gas.  In Figure~\ref{fig6}, we show the mean values
and scatter (r.m.s. values) in the real space gas--overdensity and
temperature fields, corresponding to the flux voids of the previous
sections. We show the fields both without (diamond symbols) and with
(triangles) peculiar velocities, which alter the flux distribution
observed in velocity space.  From the plot, it is clear that selecting
regions above mean flux at $z \sim 2$, as we did in the previous
Sections, allows us to identify and measure (with a large error bar)
the density in 1D gas voids with sizes larger than $\sim
10\,h^{-1}$\,Mpc in the case with peculiar velocities. If peculiar
velocities are neglected the scatter is much smaller (as expected),
and the density can be measured for somewhat smaller voids (sizes
larger than $\sim 7\,h^{-1}$\,Mpc). For even smaller voids only upper
limits of the order a few times mean density can be set, and very
little can be inferred about their temperature. The gas temperature
inside the largest flux voids is of the order of $10^4$\,K. These
regions are thus cold and, as expected from the nature of \lya forest
absorptions, sample the IGM `equation of state' around mean
density. Therefore, even when peculiar velocities are taken into
account, the physical properties of the largest voids can be reliably
studied. Furthermore, note that the effect of the peculiar velocities
for the recovered mean density and temperature inside voids is
systematically slightly higher than the corresponding values for the
case when peculiar velocities are neglected. As before, this effect is
expected since the peculiar velocity shift can move some denser clumps
of gas into voids regions that increase their overall mean density.
We stress that the 'gas voids' defined in this Section, as identified
from the transmitted flux, do not correspond to underdense regions in
the gas distribution but to regions around the mean density.

In Fig. \ref{fig1D_2} we plot the voids' densities for voids whose
sizes are larger than 7$\,h^{-1}$\,Mpc, with and without peculiar
velocities and with a varying flux threshold. The scatter is large,
but for voids regions above the mean flux level in 68$\%$ of all cases
these regions have overdensities in the range $\delta=[-0.8,0.8]$.  If
the flux threshold is lowered, large voids regions start to sample
environments at higher overdensities. The dashed line represents a
linear fit to the points (with peculiar velocities): $\delta+1
\equiv\rho/<\rho>= -0.9\,F/<F>+1.9$. This relation should describe the
mean 1D density of large ($> 7$ comoving Mpc$/h$) voids identified in
the flux distribution at $z \sim 2$, as derived from high-resolution
\lya forest QSOs spectra. We note that a systematic trend is present
and the recovered gas density is larger in the case with peculiar
velocities: we attribute this to the fact that with peculiar
velocities the corresponding gas absorptions are shifted and a void
region identified in the flux is not simply related to an underdense
region but can incorporate regions at higher density that will
increase the recovered global density.

\subsection{From 1D gas density to the 3D dark matter distribution}
\label{3d}
In this Section, we link the 1D properties of flux voids to the 3D
dark matter properties. We use a void--finding algorithm for the dark
matter distribution of the reference simulation. The algorithm is
described in more detail in \cite{colberg}. Its main features can be
summarized as follows. The starting point for the void finder is the
adaptively smoothed distribution of the matter distribution in the
simulation.  Proto--voids are constructed as spheres whose average
overdensity is below a threshold value $\delta_t$, which is a free
parameter. These proto--voids are centered on local minima in the
density field.  Proto--voids are then merged according to a set of
criteria, which allow for the construction of voids that can have any
shape, as long as two large regions are not connected by a thin tunnel
(which would make the final void look like a dumbell). The voids thus
can have arbitrary shapes, but they typically look like lumpy
potatoes.  For each void, we compute the radius $R_{DM}$ of a sphere
whose volume equals that of the void. While strictly speaking $R_{DM}$
is not the actual radius of the void, it provides a useful measure of
the size of the void.  In this work here, we adopt a value of
$\delta_t=-0.5$ (for a justification of this choice see below). As a
first, preliminary check, we compared the void size distribution of
the fiducial simulation with earlier results by \cite{colberg} and
found good agreement. Note that the size of the grid used for the
density field limits the sizes of the smallest voids that can be found
with this void finder -- in our case 1$\,h^{-1}$\,Mpc. However, as
discussed above, peculiar velocities make studies of voids smaller
than 7$\,h^{-1}$\,Mpc very hard, so we need not be concerned about the
smallest voids at all.

\begin{figure*}
\begin{center}
\includegraphics[width=18cm, height=8cm]{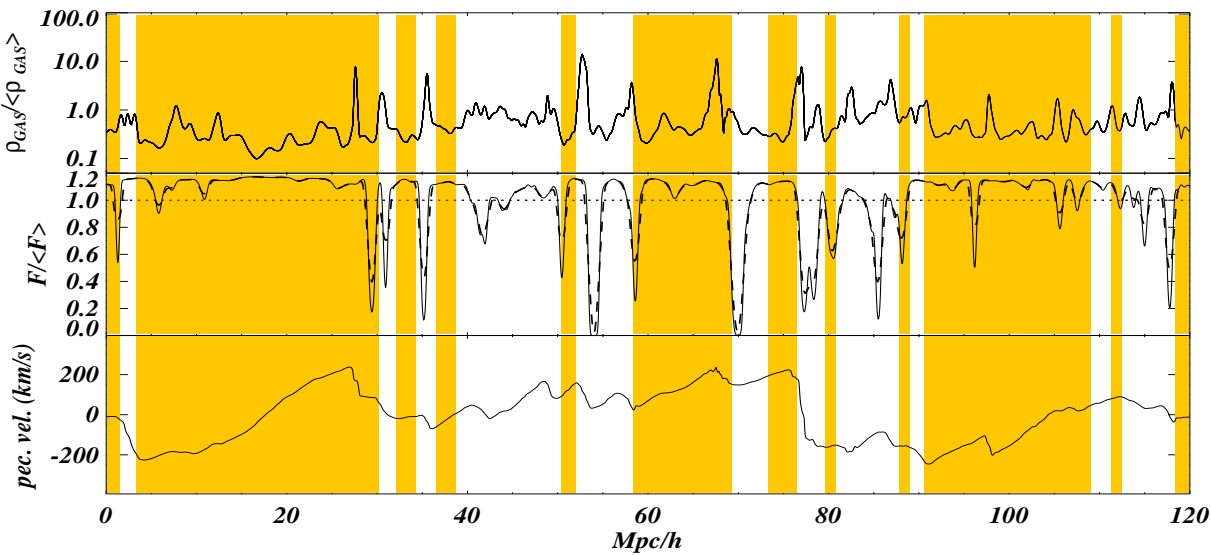}
\end{center}
\caption{Relation between 3D dark matter voids and physical quantities
  along the line--of--sight to distant QSOs. The intersections of 3D
  real--space voids in the dark matter distribution along the
  line--of--sight are represented by shaded rectangles. The top panel
  shows the 1D real--space gas overdensity, the middle one the
  simulated \lya flux in redshift space, the bottom panel the peculiar
  velocity field in km/s. Absorption features in the flux distribution
  correspond to overdensity peaks shifted by the peculiar velocity of
  the bottom panel (100 km/s roughly corresponds to 1\,$h^{-1}$\,Mpc
  comoving). The dotted line in the flux distribution sets the
  criterion for the void definition, while the dashed one represents
  the flux smoothed over 1\,$h^{-1}$\,Mpc. This particular
  line--of--sight is taken along the x-axis and at a y-coordinate of 93
  Mpc$/h$ in the slice of Figure \ref{slice}.}
\label{los11}
\end{figure*}

As a qualitative example, in Figure \ref{los11} we show one
line--of--sight through the simulated volume. We plot the 1D gas density
field in real space (top panel), the simulated flux (middle panel) in
redshift space, and the gas peculiar velocity field in km/s (bottom
panel) in real space. The $x$--axis indicates the comoving (real space)
coordinate along the line--of--sight. Flux absorption is usually
produced by gas overdensities shifted by the peculiar velocity
associated to the same gas element. For example, the absorption in the 
flux at 70\,Mpc/$h$ is produced by the gas density peak at around 
68\,Mpc$/h$, shifted by $\sim 200$ km/s -- which corresponds to roughly 
2\,$h^{-1}$\,Mpc.  

The shaded regions show the intersections with the line--of--sight of
3D voids in the dark matter distribution identified with the void
finder algorithm described above. In the middle panel we overplot the
mean flux level as a dotted line, with the flux smoothed on a scale of
1\,$h^{-1}$\,Mpc, as in Section~\ref{1Dsect}. We find that the choice
of threshold overdensity $\delta_{t}=-0.5$ for the 3D void finder
results in a good correspondence of the resulting voids with voids
defined in the flux distribution. From the bottom panel of
Figure~\ref{los11} it is clear that the void regions are expanding,
and the gas velocity fields for the largest voids are of the order of
$\pm$ 200 km/s (see the void centered on around 20\,Mpc$/h$). These
high redshift voids in the IGM will keep growing, get emptier of
matter and could be the progenitors of lower redshifts voids of sizes
similar to the Tully void (\cite{tully}).  The peculiar velocities
along the line--of--sight rise smoothly from negative values to
positive ones and are sandwiched by regions in which the peculiar
velocity shows a negative gradient, which could be the signature of a
moderate shock. It is worth emphasizing that these velocity profiles
are not always completely symmetric because the line--of--sight does
not necessarily pierce the 3D voids in their centers. Moreover, from
the top panel it is clear that the 1D density profile inside the void
is quite complex, showing small density peaks, which correspond to
small haloes that most likely host the observed void galaxy
population.

\begin{figure}
\begin{center}
\includegraphics[width=9.5cm, height=11cm]{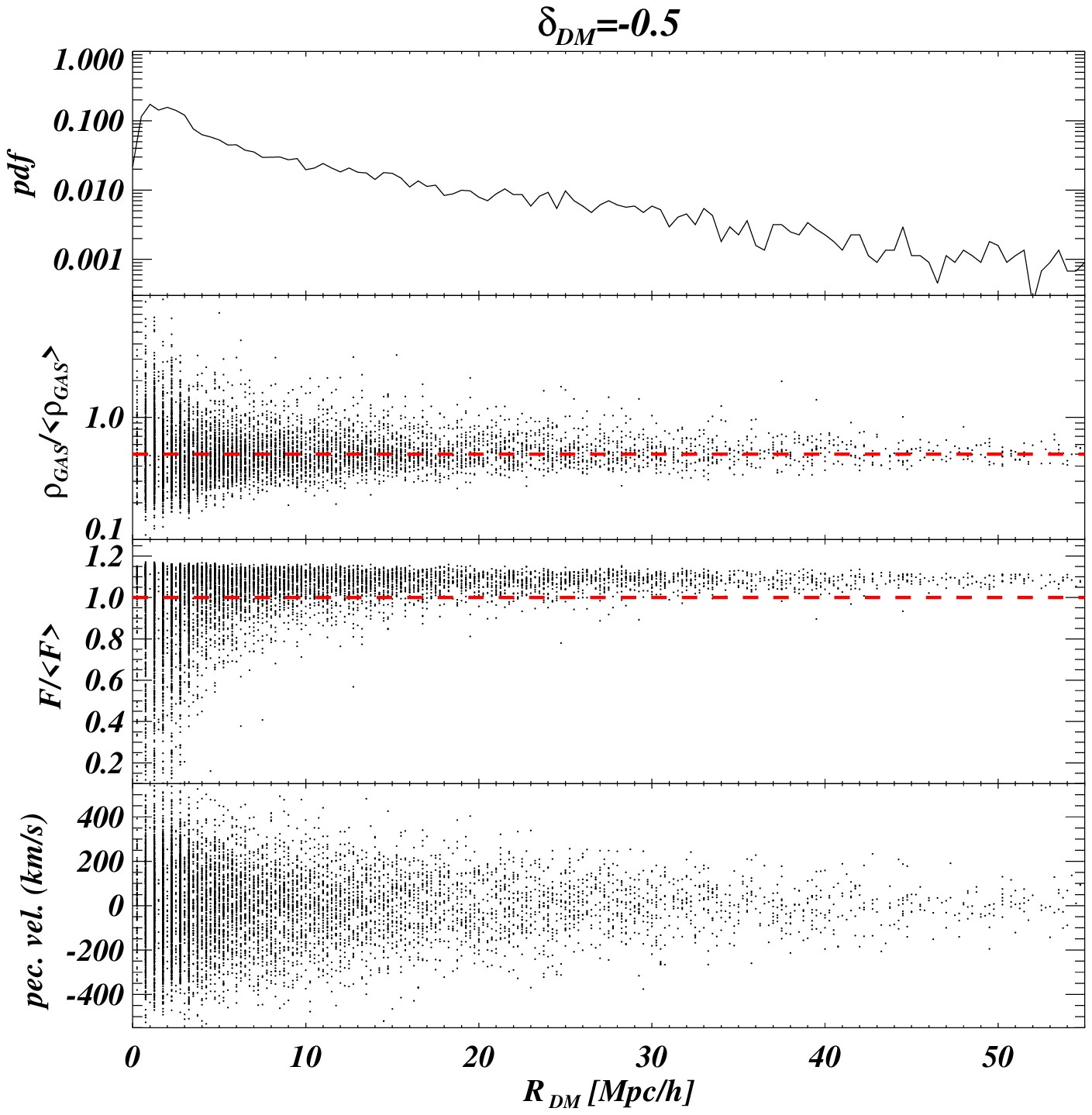}
\end{center}
\caption{Voids in the 3D dark matter distribution have been identified
with the algorithm described in the text using a density threshold of
$\delta_{DM}=-0.5$. Here we plot the 1D void flux pdf, the 1D gas median
overdensity along the line--of--sight for each identified void in the 3D
dark matter distribution, the median $F/<F>$ value in the void and the
median peculiar velocity for the same void region (from top to bottom).}
\label{figfinal}
\end{figure}

In order to provide a more quantitative picture, in Figure~\ref{figfinal} 
we plot the underlying mean values of 1D quantities obtained once the 
3D voids have been identified (using $\delta_{t}=-0.5$). The top panel 
represents the probability distribution function of the sizes of such 
1D flux voids, while in the other three panels each point shows, from 
top to bottom, the mean gas overdensity, flux and peculiar velocity field 
for a given 3D DM void.  Note that while in Section~\ref{1Dsect} we were
interested in measuring the properties of voids once the mean flux
threshold was set as a criterion to define voids, here we use the 3D
DM distribution to look for the corresponding 1D (\lya forest)
flux--related quantities.

The overall picture suggests that large voids ($>7-10$\,Mpc$/h$) are
usually related to flux values above the mean, and the corresponding
mean values of gas density and peculiar velocities have less scatter
than smaller voids. The corresponding mean value for the gas
overdensity is $\delta_{\mbox{GAS}}=-0.43$, so the 3D void region is
slightly denser in gas than in dark matter, something that can be
expected since baryons feel pressure and are thus more diffused than
the collisionless dark matter (e.g. \cite{viel02}). The different
value of $\delta_{\mbox{GAS}}$ than the one found in
Section~\ref{1Dflux} does not contradict our earlier findings, since
now the requirement to define a void is {\it less} stringent. Now, the
actual void has to be a quasi--spherical 3D region in the DM
distribution and not a connected 1D region with flux above the mean
flux.  This means that in 1D one is more sensitive to clumps of matter
producing absorptions, while the same clump will have a smaller impact
on the mean density of the 3D region surrounding it.  Cases for which
the flux voids correspond to actual DM voids are avaraged out
statistically by small lumps of gas around the mean density along the
line--of--sight, that cause absorption and do not alter the global 3D
DM density of the void region

For this population of large voids the typical scatter in the peculiar
velocity field is 170\,km/s.  We thus have shown that the population
of 1D large flux voids, i.e. regions above the mean flux level, as
estimated from a set of mock high resolution spectra at $z\sim 2$,
traces reasonably faithfully a population of 3D dark matter voids of
similar sizes, with a typical $\delta_{t}=-0.5$.

We have also checked that the scatter plots of Figure~\ref{figfinal} 
are not very sensitive to the cosmological model (i.e. different values
of $\sigma_8$), which means that the voids physical properties are the 
same for different values of the power spectrum amplitude.

\section{Conclusions and perspectives}
\label{conclusions}
We used \lya forest QSO spectra to constrain the void population at 
$z\sim 2$ at a range of scales and redshifts, which cannot be probed by 
other observables.  The main conclusions can be summarized as follows:
\begin{itemize}
\item
the properties (sizes) of voids identified in simulations from  the flux 
distribution as connected regions above the mean flux level are in 
good agreement with observed ones;
\item
using a set of hydrodynamical simulations and varying all the
astrophysical and cosmological parameters, along with checking
box size and resolution effects, we find that the flux void size
distribution is a robust statistics that depends primarily on the flux
threshold chosen to define the voids;
\item
the flux seems to be a reliable tracer of the underlying dark matter
distribution, altering the statistical properties of the dark matter
density field (changing amplitude, shape, adding or suppressing power
in a wavenumber dependent manner) changes the properties of the void
population although at present is difficult to give constraints on
cosmological parameters using these statistics;
\item
flux voids, i.e. connected regions above the mean flux levels,
correspond to gas densities around the mean density, regardless of the
role of peculiar velocities, which contribute to a scatter in this
relation, but do not alter significantly the mean values;
\item
linking 1D voids to the corresponding 3D dark matter voids is more
difficult, and we used a void--finding algorithm for that: we find 
that 3D DM voids with a mean overdensity of $\delta_{t}=-0.5$
correspond to the flux voids that we have defined from the QSO
spectra. However, the correspondence is good only for voids with 
sizes larger than about $7-10$\,Mpc$/h$.
\end{itemize}

In this paper, we made a preliminary attempt to link the population of
voids in the transmitted \lya flux to the underlying 1D gas density
and temperature and 3D dark matter density. The use of \lya
high--resolution spectra is important in the sense that it explores a new
regime in scales, redshifts and densities which is currently not
probed by other observables: the scales are of order few to tens of
Mpc, the redshift range is between $z=2-4$, while the densities are
around the mean density.  Further studies, that could possibly rely on
wider data sets such as the low--resolution SDSS data (\cite{shang}),
on other future observables at higher redshifts (\cite{aloisio}) or on
tomographic studies involving QSO pairs or multiple line--of--sights
(\cite{dodorico}, \cite{saitta07}) would be important in understanding
the dynamical, thermal and chemical evolution of the void population
and the interplay between galaxy and the IGM over a large fraction of
the Hubble time.

\section*{Acknowledgments.}
Numerical computations were done on the COSMOS supercomputer at DAMTP
and at High Performance Computer Cluster (HPCF) in Cambridge (UK).
COSMOS is a UK-CCC facility which is supported by HEFCE, PPARC and
Silicon Graphics/Cray Research. The authors thank the Virgo Consortium
and the Lorentz Center in Leiden for their hospitality during the
Virgo Workshop in early 2007, where this work was begun.  We thank the
anonymous referee for his/her very careful reading of the paper and
the comments made.

\bibliographystyle{mn2e} \bibliography{master2.bib}

\end{document}